\documentstyle[bo99B,epsfig]{article}

\def\etal{et al.}

\title{Systematic effects in the interpretations of 
          Cluster X-ray Temperature functions}

\author{S.~M.~Molnar $^{1,2}$ and K.~Jahoda $^1$}                         

\affil{1) Laboratory for High Energy Astrophysics, Code 662
          Goddard Space Flight Center
          Greenbelt, MD 20771 \\
       2) NAS/NRC Research Associate}

\begin{document}

\maketitle

\begin{abstract}
The formation and evolution of clusters of galaxies are sensitive to 
the underlying cosmological model. Constraints on cosmological 
parameters of cold dark matter models have been derived from mass, 
temperature and luminosity functions of clusters.
We study the importance of including cluster formation history
and a correction for collapsed fraction of objects
in determining the cluster X-ray temperature function.
We find that both effects are important.
We compare temperature functions obtained by using a 
power law approximation for the mass variance normalized to X-ray clusters
to those obtained by using a COBE normalized full CDM treatment.
We conclude that the temperature function could be a powerful test
on the average density of the Universe if we could
find the correct way of interpreting the data.

\end{abstract}

\section{Introduction}

Clusters of galaxies are the largest known virialized gravitational
systems in the Universe. According to our working hypothesis of structure
formation, the cold dark matter (CDM) model, clusters form via gravitational 
collapse of high density peaks of primordial density fluctuations.
The distribution and evolution of these peaks strongly depend on the 
CDM model parameters. We use the following parameterization for our CDM models:
$\Omega_0$, $\Omega_b$, $\Omega_\Lambda$, $h$ (matter density,
baryon density, cosmological constant, all in units of the
critical density, all at $z$ = 0, and $h$ is the Hubble constant in units of 
$100 \, \rm Km\,s^{-1}\,Mpc^{-1}$).
Semi-analytical methods, like the Press-Schechter approximation, use some version of 
a spherical collapse model to predict the distribution of collapsed objects and 
their masses as a function of redshift.
Assuming a physical model for the clusters and their intracluster medium,
one can translate cluster virial masses to X-ray luminosity, or temperature of
the intracluster gas, which can be compared to observations.
We use the temperature function of clusters in our study.
The most popular method is to assume a power law approximation for the
power spectra of CDM models, use the Press-Schechter mass function (PSMF) to predict
the abundance, and then derive the virial temperature to get the intracluster gas temperature,
or integrate over formation epochs to determine the mass-temperature conversion.
This method has the power spectrum normalization and the power law exponent of the
mass variance on cluster scales as free parameters and determines matter density,
and has no dependence on the Hubble constant.
We use COBE normalized spectra of CDM models to derive the mass variance and the 
Press-Schechter mass function to derive the distribution of cluster masses.
This method returns the matter density as a function of the Hubble constant,
thus one can not derive either parameter separately.
We use four different methods to obtain theoretical temperature functions.
We integrate over formation epochs in methods A and B but not in methods C and D.
We take into account the collapsed fraction of objects in methods A and C and
not in B and D. Method A thus take both effects and method D none of them
into account. For each method (A, B, C, and D) we use a grid of CDM model
parameters, $0.2 < \Omega_m < 0.9$ and $0.2 < h < 0.9$,
and calculate the temperature functions assuming open and flat CDM cosmologies.
We compare these theoretical temperature functions to data and determine the best fit models
by minimizing the corresponding $\chi^2$.

\section{Outline of the methods}

The power law approximation for the mass variance 
assumes a power exponent $\alpha$, and derives the mass variance as
$\sigma(M) = \sigma_8 (M/M_{15})^{-\alpha}$, 
where $M_{15} = M /(10^{15} M_\odot)$, and $M$ is the mass we will identify 
with the virial mass in the PSMF.
It is commonly assumed that the power spectrum then can be approximated by a power law
with an exponent $\alpha = (n_{PS}+3)/6$, but that is true only 
if the power spectrum could be approximated by one power law
in all scales (note the integral in equation 1). 
That is not true for CDM models, therefore we quote the mass variance power exponents 
on cluster scale for power law approximations.
Instead of the approximation we use COBE normalized power spectra of CDM models 
(Hu and Sugiyama 1996) and obtain the mass variance from a numerical integral

\begin{equation}
 \sigma^2_{R(M)} = (1 / 2 \pi^2) \int P(k) W^2(kR) k^2 dk
,
\end{equation}
where $P(k)$, $W(kR)$, $R(M)$, and $k$ are the power spectrum, filter function
in Fourier space, the radius of filtering, and the co-moving wavenumber.
We use the standard PSMF corrected for collapsed fraction $f_c$, 
which gives the fraction of matter in collapsed objects:
$n(z, M)\, dM = (f_c/f_c^{PS}) \; n_{PS}(z, M)\, dM$,
where $f_c^{PS}$ is the Press-Schechter collapsed fraction (Martel and Shapiro 1999).
In order to obtain the temperature function, we
assume that clusters form over a period of time and integrate over formation
$n_T(z, M) = \int_z^\infty F(M, z_f, z) {dM \over dT}(T, z_f, z) d z_f$, 
where $F(M, z_f, z)$ depends on $\sigma(M)$, $\delta_c(z)$, and $n(z_f, M)$
(Kitayama and Suto 1996).
We used the spherical collapse model virial temperature (Eke et al. 1998):

\begin{equation}
  M(T) = M_{15} \Bigg[ {\beta k \,T[keV] \over 1.38 keV (z+1) } \;\; \Bigg]^{3 \over 2} 
                \Bigg[ {\Omega(z) \over \Omega(0) \Delta_c(z) } \;\; \Bigg]^{1 \over 2}
.
\end{equation}
We use $\beta =1$ as suggested by numerical simulations (Eke et al. 1998).

\section{Results and Conclusions}

In Figure~1a we show the resulting best fit temperature functions using the power 
law approximation of Donahue and Voit (1999) and Blanchard et al. (1999) as we 
reconstructed them following their method.
We used Horner et al. (1999)'s compilation of data 
taken from Edge et al. (1990) Henry and Arnaud (1991), and Markevitch (1998) 
(squares, diamonds, and triangles).
The solid and dashed dotted lines represent best fit flat model,
$\Omega_m = 0.27$, $\sigma_8 = 0.73$ $\alpha = 0.13$, of
Donahue and Voit (they integrated over cluster formation epoch) and their
best fit model, but not integrated over cluster formation epoch.  
Blanchard et al.'s best fit flat model, 
$\Omega_m = 0.735$, $\sigma_c = 0.623$, $\alpha = 0.18$, is represented by a long dashed line. 
The short dashed line represents Blanchard et al.'s best fit temperature function, but
integrated over cluster formation epoch. 
Open model temperature functions (not shown) are very similar to those of the flat models.
The difference between temperature functions from methods with and without integration over 
cluster formation epoch is significant even with existing data.
The large difference between density parameters derived by Donahue and Voit, and 
Blanchard et al. is due to the different normalizations of the $M(T)$ relation and not 
integration over cluster formation, which can cause a change less than 0.1 in the 
density parameter (cf. next paragraph).

Figure~1b shows temperature functions of the best fit flat CDM model 
(best fit using method A) with $\Omega_m = 0.39$, $\Lambda = 0.61$, $h = 0.5$, and this 
best fit model, but temperature functions with and without taking the collapsed fraction 
and/or integrating over cluster formation epochs (our methods B, C and D). 
The data are the same as in figure 1a. 
The solid, long dashed, short dashed and dashed dot lines
represent temperature functions using our methods A, B, C and D.
Open models behave similarly so we do not show them here.
The smallest effect is the correction for collapsed fraction of objects
which is about the size of the error bars of the data, thus an important effect 
(compare methods A and B).
A larger difference is caused by integration over formation epoch (methods A and C).
If one fits models without integrating and/or taking collapsed fraction into account 
(methods B, C and D), one get matter densities 
$\Omega_m = 0.36$, $\Omega_m = 0.34$, and $\Omega_m = 0.31$ (assuming $h = 0.5$).
Not taking either effects into account causes about 0.1 change 
in the derived matter density (compare results using methods A and D). 
As we can see, the temperature function is very sensitive to CDM model parameters 
if one uses a full CDM treatment, and thus a precise determination of the matter density 
is possible if the Hubble constant is known. We should keep in mind, however, that the
normalization of the $M(T)$ relation, for example, causes much larger error  
in determining $\Omega_m$.
Since this method does not allow us to separate the density parameter from 
the Hubble constant, the result is a best fit function of the two.
We can make use of the results from the power law approximation
which gives the best fit density parameter, and check if the corresponding
Hubble constant is reasonable using our methods.
We find that our best fit CDM models yield the same matter density as the 
best fit models of Donahue and Voit (1999) and Blanchard et al. (1999)
if we use $h = 0.55$ and $h = 0.35$. Thus Blanchard et al.'s method is 
only marginally compatible to ours.

Figure 1b shows that  systematic errors from interpretation of
the temperature function are larger than the error bars on even the existing data.
We conclude that, if the Hubble constant is known, a comparison between the
observationally derived cluster temperature function and those
derived from a full CDM treatment may yield an accurate determination
of the density parameter (with an error less then 0.05), however, this can be done 
only if we find other ways to derive a correct theoretical model to interpret the data.

\begin{figure}
\centerline{\psfig{file=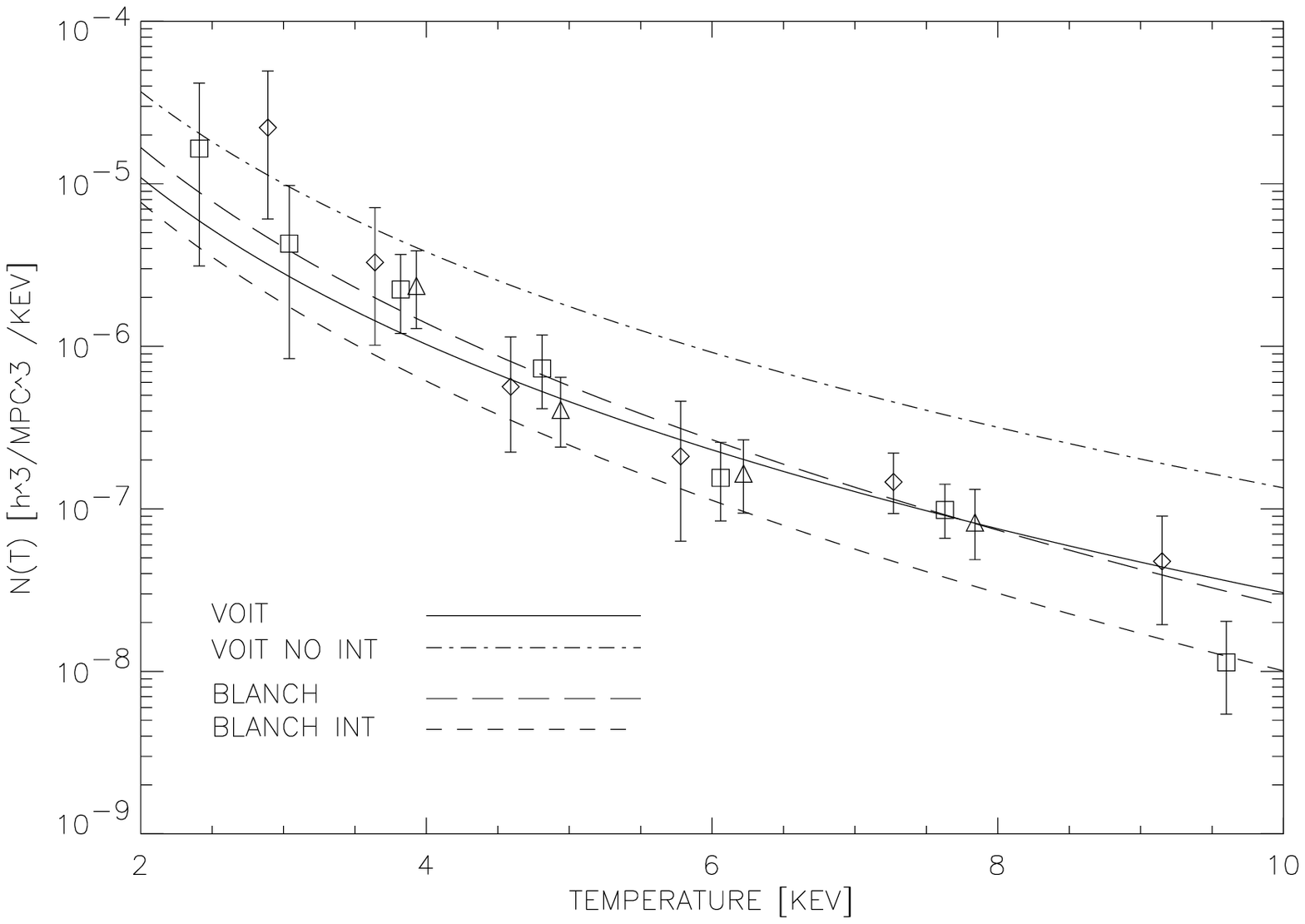, width=6cm}
            \psfig{file=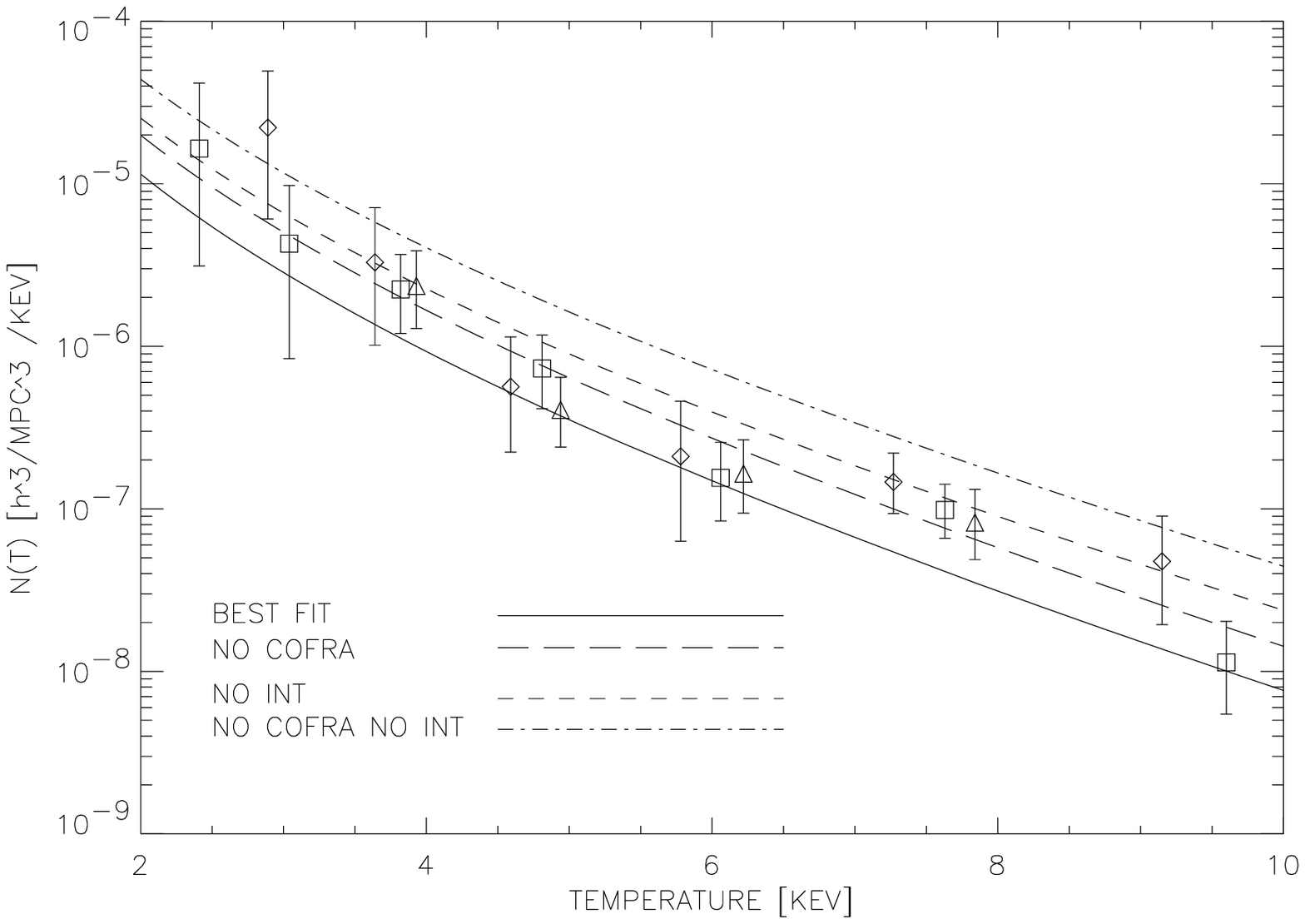, width=6cm}}
\caption[]{a) left panel: temperature functions using best fit power law approximations;
           b) right panel: temperature functions of the best fit CDM flat model using
           different methods. See text for details.}
\end{figure}

\begin{acknowledgements}
We thank N. Aghanim and R. Mushotzky for valuable 
discussions, and Don Horner for providing us his compilation of the temperature data.
We thank N. Aghanim for her hospitality at the Institute of Astronomy of Orsay University.
\end{acknowledgements}

\end{document}